\documentclass[]{aastex631}
\usepackage{amsmath}

\newcommand{\tAA}[0]{\text{\AA}}

\begin{document}

\title{Extract cleaned Swift/UVOT UV grism spectra with {\tt uvotpy} package}

\author[0000-0003-2915-7434]{Hao Zhou}
\affiliation{Key Laboratory of Dark Matter and Space Astronomy, Purple Mountain Observatory, \\
Chinese Academy of Sciences, Nanjing 210023, China}

\author[0000-0001-9078-5507]{Stefano Covino}
\affiliation{INAF/Brera Astronomical Observatory, via Bianchi 46, I-23807 Merate (LC), Italy}
\affiliation{Como Lake centre for AstroPhysics (CLAP), DiSAT, Università dell’Insubria, via Valleggio 11, I-22100 Como, Italy}

\author[0000-0003-4977-9724]{Zhi-Ping Jin}
\affiliation{Key Laboratory of Dark Matter and Space Astronomy, Purple Mountain Observatory, \\
Chinese Academy of Sciences, Nanjing 210023, China}
\affiliation{School of Astronomy and Space Science, University of Science and Technology of China, \\
Hefei 230026, China}

\author[0000-0002-8966-6911]{Yi-Zhong Fan}
\affiliation{Key Laboratory of Dark Matter and Space Astronomy, Purple Mountain Observatory, \\
Chinese Academy of Sciences, Nanjing 210023, China}
\affiliation{School of Astronomy and Space Science, University of Science and Technology of China, \\
Hefei 230026, China}

\author[0000-0002-9758-5476]{Da-Ming Wei}
\affiliation{Key Laboratory of Dark Matter and Space Astronomy, Purple Mountain Observatory, \\
Chinese Academy of Sciences, Nanjing 210023, China}
\affiliation{School of Astronomy and Space Science, University of Science and Technology of China, \\
Hefei 230026, China}

\author[0000-0003-4650-4186] {N. Paul Kuin} \affiliation{University College London, Mullard Space Science Laboratory, Holmbury House, Dorking, Surrey, UK}

\correspondingauthor{Hao Zhou, Zhi-Ping Jin}
\email{haozhou@pmo.ac.cn, jin@pmo.ac.cn}

\begin{abstract}
The ultraviolet/optical telescope (UVOT) onboard the Neil Gehrels Swift Observatory is capable of imaging with 7 lenticular filters and of taking slitless spectra with 2 grisms. Both image and grism data have been widely used to study gamma-ray bursts, supernovae and other ultraviolet/optical transients, and proved UVOT is a powerful instrument in time-domain astronomy. However, the second order contamination, for blue sources, strongly limits the red end of ultraviolet (UV) grism spectra. This, in turn, reduces the valid wavelength range to only about 33\,\% of the total. However, to explore the broadband spectral energy distribution of GRBs at the early stage, a larger valid wavelength range is required. Hence based on the {\tt uvotpy}\footnote{\url{https://github.com/PaulKuin/uvotpy}} package \citep{2014ascl.soft10004K,2015MNRAS.449.2514K}, we propose a method to remove the second order contamination from UV grism spectra (nominal mode) up to $\sim4000\,\tAA$, i.e., about 70\,\% of the full wavelength range. The 1-$\sigma$ systematic uncertainty of this method is $\sim11.2\,\%$. In addition, if a source is red enough, the red end of the valid range could reach $\sim5000\,\tAA$. The source code is available on GitHub\footnote{\url{https://github.com/HaoZhou0810/cluvotpy}}.
\end{abstract}

\keywords{Ultraviolet astronomy (1736), Astronomy data reduction (1861)}

\section{Introduction} \label{sec:intro}

The Neil Gehrels Swift Observatory \citep{2004ApJ...611.1005G} was launched in 2004 with the primary purpose of exploring the most powerful explosions in the universe, i.e., gamma-ray bursts (GRBs), and it was designed to be capable of performing rapid follow-up observations for burst events within minutes. There are three instruments on the satellite that cover the $\gamma$ -ray to optical bands: the burst alert telescope \citep[BAT, ][]{2005SSRv..120..143B}, the X-ray telescope \citep{2005SSRv..120..165B} and the ultraviolet/optical telescope \citep{2004SPIE.5165..277M,2005SSRv..120...95R}. UVOT is equipped with 7 lenticular filters to obtain broadband photometry \citep{2008MNRAS.383..627P,2011AIPC.1358..373B} and 2 grisms (UV and visible) to take slitless spectra \citep{2015MNRAS.449.2514K}. UVOT has a photon counting detector and is capable of recording the arrival time and the position of each incident photon. In the past 2 decades, UVOT has proved to be a powerful instrument to study UV/optical counterparts of GRBs, for example, the onset of the afterglow in the GRB 081203A \citep{2009MNRAS.395L..21K}, optical prompt emissions of GRB 110205A \citep{2016ApJ...831L...8G} and GRB 241030A \citep{2024GCN.38055....1W}, and the brightest UV/optical flare in GRB 220101A \citep{2023NatAs...7.1108J}.
%, the bright ordinary GRB 130427A \citep{2014Sci...343...48M}, the famous ``naked eye" burst GRB 080319B \citep{2008Natur.455..183R,2009ApJ...691..723B}, 2 long-short GRBs: GRB 060505 and GRB 060614 \citep{%2006Natur.444.1044G,
%2006Natur.444.1053G,2007ApJ...662.1129O,2021arXiv210907694J,2007A&A...470..105M},
%and the double peaks of the optical emission of GRB 241030A \citep{
%2016ApJ...831L...8G,
%2024GCN.38055....1W}. 
In addition, research on the further development of UVOT capabilities, for example, saturation correction \citep{2013MNRAS.436.1684P,2023ApJS..268...65Z}, is continuing.

After the full calibration of UVOT grisms in 2015, UVOT grism data has been widely used to study supernovae %\citep{2015ApJ...805...74B,2015ApJ...810...86C,2015ApJ...813...30S,2018MNRAS.480..572A,2018MNRAS.479..517P,2018ApJ...858...78G,2018MNRAS.474.2679A,2017MNRAS.464.3281W,2019MNRAS.485.5120B,2020MNRAS.491.5897P,2019ApJ...887..169H,2022MNRAS.513.5642B,2021MNRAS.501.1951C,2020MNRAS.499.4814P,2020MNRAS.492.3107L,2022ApJ...934..134V}.
\citep{2015ApJ...805...74B,2015ApJ...813...30S,2018MNRAS.479..517P,2019MNRAS.485.5120B,2020MNRAS.491.5897P,2019ApJ...887..169H,2022ApJ...934..134V}.
Since the UV grism is not blazed, for some blue sources, the second order intensity where the first order wavelength $\gtrsim3000\,\tAA$ could be comparable with the first order intensity. Although the wavelength range of UV grism spectra is from 1700 to 5000,\tAA \citep{2015MNRAS.449.2514K}, the valid red end is limited by  contamination: only data with $\lambda\lesssim3000\,\tAA$ is reliable \citep{2019MNRAS.487.2505K,2020MNRAS.491..655K}. Therefore, when analyzing the continuum, only $\lesssim40\,\%$ of the data is usable. If the source is red enough, the second order contamination is negligible (e.g., for most supernova mentioned above, their temperatures are less than $10^4\,$K or the extinction is large).

For BAT triggered GRBs, the automatic UVOT follow-up observations may contain a 50 seconds UV grism exposure in the nominal mode, which typically begins $\sim250\,$s after the trigger. The source position is usually near the default position (i.e., close to the center of the UVOT detector), and at the default position, the second order contamination typically starts from $\sim3000\,\tAA$. As a result, the first order UV grism spectra of GRBs were only used to confirm redshifts in the literature, e.g., GRB 081203A \citep{2009MNRAS.395L..21K} and GRB 130427A \citep{2014Sci...343...48M}. Simultaneous X-ray and UV grism observations could indeed provide useful data to study broad-band spectral energy distributions (SEDs) of GRBs at the early stage, which could provide some clues to reveal the physics of GRBs. However, the wavelength range of the first order free of the second order contamination is too short to get robust conclusions or constraints. Hence, to build reliable broadband SEDs (from UV/optical to X-ray/$\gamma$-ray) of GRBs at the early stage, we propose here a method to remove the second order contamination from UV nominal spectra (i.e., taken with the UV grism in the nominal mode) up to $\sim4000\,\tAA$, which is also helpful for blue transients, e.g., AT2018cow \citep[a fast blue optical transient,][]{2019MNRAS.487.2505K}. The 1-$\sigma$ systematic uncertainty of this method is $\sim11.2\,\%$. The red end of the valid wavelength, $\sim4000\,\tAA$ is limited by the third order contamination, and when the source is red enough that the third order contamination is negligible, the red end could reach $\sim5000\,\tAA$, as discussed in Section \ref{sec:demonstration}. For UV spectra taken in the clocked mode, there are some problems about the flux calibration, so we did not calibrate them in this paper.

The principle of the method is described in Section \ref{sec:principle}. Section \ref{sec:cal} introduces the calibration and the systematic uncertainty of this method. Section \ref{sec:demonstration} presents the comparison of several cleaned spectra derived by the method with reference spectra, and the look-up table for the quick estimation of the degree of the second-order contamination. Some instructions and the work flow of this method are summarized in Section \ref{sec:con}.

\section{The principle of the Clean Extraction} \label{sec:principle}
Figure \ref{fig:2dspec} shows a UVOT nominal 2d spectrum of a bright white dwarf AG+81 266. The 2d spectrum suffers from strong coincidence loss and is also saturated, hence the overlap of the first and second order spectra for $\gtrsim3000\,\tAA$ is clearly seen. As a result, the red end of the valid wavelength range is limited by the second order contamination. In addition, the overlap of the first and the third order is also seen for $\gtrsim4200\,\tAA$, which constrains the valid range up to $\sim4000\,\tAA$ after the removal of the second order contamination.
%As a result, the red end of the valid wavelength range is limited by the second order contamination.
The spatial separation between different orders depends on the source positions on the detector, and using a narrow extraction aperture one can avoid contamination of higher orders to larger wavelength. Hence, the second order contamination not only depends on the observation set-up, but also on the extraction configuration, which means there are strict requirements to be satisfied to remove the second order contamination (summarized in Section \ref{sec:con}).

The high-order overlapping is also seen in the HST WFC3 G280 grism data. \cite{2017wfc..rept...20P} determined the trace and the wavelength calibration of the first order spectrum at the center of the detector. In 2020, with about 600 observations, a comprehensive determination of traces, wavelength calibration, sensitivity calibration for different orders across the entire field of view of the detector was carried out\footnote{\url{https://www.stsci.edu/files/live/sites/www/files/home/hst/instrumentation/wfc3/documentation/instrument-science-reports-isrs/_documents/2020/WFC3-ISR-2020-09.pdf}}. Hence, with the first order data free of the high order contamination, the contribution of high orders to the photon count in the extraction aperture can be estimated and removed properly. In principle, the method proposed in this article is a simplified version of the one applied to HST WFC3 G280 grism data: we focus on a region near the default position of the UVOT detector and remove the contribution of the second order from the total count, which is extracted with the optimal/standard aperture defined in \citet{2015MNRAS.449.2514K}.

\subsection{Definitions of parameters}
We use the same parameter set to describe grism spectra as in \cite{2015MNRAS.449.2514K}. The following parameters are the most important in this work:
\begin{itemize}
    \item The anchor position of the n-th order spectrum ${\rm ANK_n}$. For the UV grism, ${\rm ANK_n}$ is equal to the detector coordinate in pixels where 2600\,\tAA~ is located.
    \item The pixel number PN. Adding a constant number to the column ID of the 2d spectrum, so that the shifted column ID of the first order anchor position in the 2d spectrum equals 0. The shifted column ID is denoted as PN.
    \item The function converting the PN to the n-th order wavelength $\lambda_{\rm n}({\rm PN})$, and the inverse function ${\rm PN}_{\rm n}(\lambda)$.
    \item The width of wavelength per pixel of the n-th order spectrum at PN, $\Delta\lambda_{\rm n}({\rm PN})$.
    \item The effective area of the n-th order at PN, ${\rm EA}_{\rm n}({\rm PN})$, when using a specified extraction aperture, e.g., the extraction aperture in Figure \ref{fig:2dspec}. For example, the second order effective area is $\sim0\,{\rm cm}^2$ at ${\rm PN}\lesssim100$, since only a very small fraction of the trace of the second order spectrum overlaps with the extraction aperture.
    \item The corrected count rate of the n-th order at PN, ${\rm CR}_{\rm n}({\rm PN})$
\end{itemize}

The conversion between the flux density and the count rate is:
\begin{equation}\label{equ:cr2flux}
    f_{\lambda,{\rm n}}({\rm PN})\equiv f_\lambda(\lambda_{\rm n}({\rm PN}))=\frac{hc~{\rm CR}_{\rm n}({\rm PN})}{\lambda_{\rm n}({\rm PN})~{\rm EA}_{\rm n}({\rm PN})~\Delta\lambda_{\rm n}({\rm PN})}.
\end{equation}
It is convenient to define the factor converting the count rate to the flux density
\begin{equation} \label{equ:def_cf}
    {\rm CF}_{\rm n}({\rm PN})=hc/\lambda_{\rm n}({\rm PN})/{\rm EA}_{\rm n}({\rm PN})/\Delta\lambda_{\rm n}({\rm PN}),
\end{equation}
and rewrite Equation \ref{equ:cr2flux} as
\begin{equation} \label{equ:cr2flux_short}
    f_{\lambda,{\rm n}}({\rm PN})={\rm CF}_{\rm n}({\rm PN})~{\rm CR}_{\rm n}({\rm PN}).
\end{equation}
Though Equation \ref{equ:cr2flux_short} is expressed in PN, it is easy to convert PN to wavelength with the function $\lambda_{\rm n}({\rm PN})$, and using PN instead of the wavelength as the independent variable makes the formula concise and easy to understand.

\subsection{Subtracting the second order}
\begin{figure}[h]
    \centering
    \includegraphics[width=0.8\linewidth]{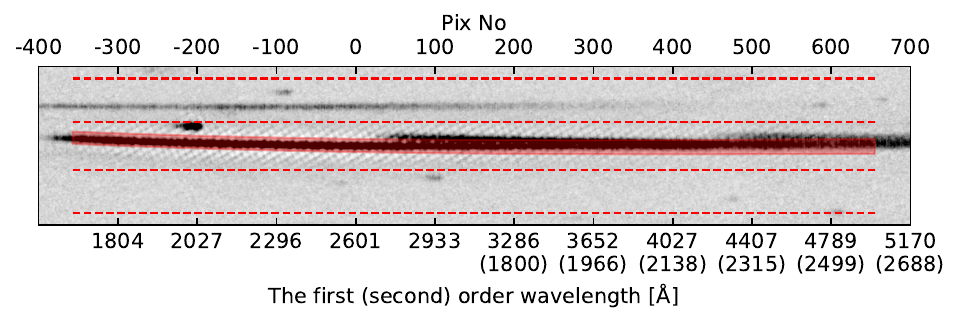}
    \caption{The 2d UV nominal spectrum of AG+81 266. The red region represents the trace of the optimal/default extraction aperture spanning from $1700\,\tAA$ to $5000\,\tAA$. Upper and lower background regions are shown by dashed red lines above and below the central spectrum, respectively. From pixel number $\sim50$ (i.e, $\sim2800\,\tAA$), the second order spectrum becomes visible, and begins to overlap with the extraction aperture. Because AG+81 266 is bright and blue enough, the third order spectrum is also visible starting from pixel number $\sim450$. Both the second and the third order wavelength at the PN where the first order wavelength is $5000\,\tAA$ (i.e., in this figure, the PN $\sim660$), are less than $2600\,\tAA$. Hence, the contamination can be removed with the first order spectrum free of the contamination ($\lambda<2600\,\tAA$).}
    \label{fig:2dspec}
\end{figure}

The total corrected count rate at PN is the composition of spectra with different orders: 
\begin{equation}
    {\rm CR}({\rm PN})=\sum_{\rm n}^{\infty}{\rm CR}_{\rm n}({\rm PN})=\sum_{\rm n}^{\infty}f_{\lambda, {\rm n}}({\rm PN})/{\rm CF}_{\rm n}({\rm PN}).
\end{equation}
For the 2d spectrum in Figure \ref{fig:2dspec}, the second order wavelength at ${\rm PN_1(5000\,\tAA)}$ is $\sim2600\,\tAA$, and for wavelength range $\lambda<2600\,\tAA$, the first order spectrum is clean. Hence, it is possible to remove the second order contamination with the first order spectrum free of the contamination. There is no wavelength calibration for the third order, so the wavelength range of the third order for $450\lesssim{\rm PN}<{\rm PN_1}(5000\,\tAA)$ is unknown. However, it is certain that $\lambda_3({\rm PN_1}(5000\,\tAA))$ is less than $2600\,\tAA$.

Neglecting the third order and higher order spectra, the corrected count rate of the cleaned first order spectrum at PN can be written as:
\begin{equation} 
\begin{aligned}
    f_{\lambda, 2}({\rm PN})&=f_\lambda(\lambda_2@{\rm PN}) \label{equ:clean1st}\\
    &=f_{\lambda, 1}({\rm PN}_1 @\lambda_2 @{\rm PN}) \\
    &={\rm CF}_1({\rm PN}_1 @\lambda_2 @{\rm PN})\times{\rm CR}_1({\rm PN}_1 @\lambda_2 @{\rm PN}), \\
    {\rm CR}_1({\rm PN})&={\rm CR}({\rm PN}) - \frac{f_{\lambda, 2}({\rm PN})}{{\rm CF}_{\rm 2}({\rm PN})},~0<{\rm PN}<650.
\end{aligned}
\end{equation}
To improve the readability, parentheses are replaced by the symbol $@$, i.e., $f@g@x=f(g(x))$. The pattern ${\rm PN}_1 @\lambda_2 @{\rm PN}$ means the pixel number where the first order wavelength equals the second order wavelength at PN. For example, it can be read from Figure \ref{fig:2dspec} that ${\rm PN}_1 @\lambda_2@200\approx-300$. Equation \ref{equ:clean1st} shows the intensity of the second order spectrum at PN can be estimated with the factor converting the count rate to the flux density and the corrected count rate of the first order spectrum.

\subsection{Calculation of the factor converting count rate to the flux density}
Equation \ref{equ:def_cf} shows that at PN, the factor converting the count rate to the flux density depends on the central wavelength, the width of wavelength and the effective area. Reliable wavelength and effective area calibrations almost across the entire detector can be found in the built-in calibration database directory of {\tt uvotpy} package. As mentioned above, the effective area depends on the extraction aperture and the source position. Hence, it is necessary to specify an extraction aperture, and the default/optimal extraction aperture in the {\tt uvotpy} package, described in \citet{2015MNRAS.449.2514K}, is a good choice. The applicable region is discussed in Section \ref{sec:cal}.

The function converting PN to wavelength of the n-th order spectrum $\lambda_{\rm n}$ is expressed in terms of polynomials, and {\it DISP*} keywords in the header of the extracted spectrum represent coefficients of polynomials. With $\lambda_{\rm n}$, it is convenient to derive $\Delta\lambda_{\rm n}({\rm PN})=\lambda_{\rm n}({\rm PN}+0.5)-\lambda_{\rm n}({\rm PN}-0.5)$ for integer PN. For the effective area, the data in the {\tt uvotpy} calibration database is sampled sparsely in the wavelength space, hence a cubic spline interpolation method is applied to derive the value at the specified wavelength.

\section{Calibration of the Clean Extraction}\label{sec:cal}
\subsection{The second order effective area}
We found that the second order effective area in the {\tt uvotpy} package is systematically larger than the true value\footnote{The second order was biased to a larger value with the thought this would ensure a good error estimate of its contribution.} when using the default/optimal extraction aperture, since the flux density of the second order is always overestimated. The method described here to correct the first order extraction for the second order will be called ``Clean Extraction". Hence, four white dwarfs are selected from \cite{2015MNRAS.449.2514K} to calibrate the second order effective area when using the default/optimal extraction aperture. The search radius is set $4\,\arcmin$, so that anchor positions will be not too far away from the default position, and a total of 46 observations are used to calibrate the second order effective area (Please refer to Table \ref{tbl:obs} for details).

\begin{figure}[h]
    \centering
    \includegraphics[width=0.49\linewidth]{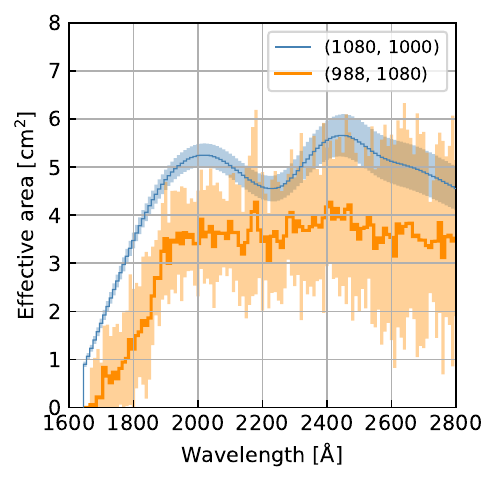}
    \caption{The effective area of the second order spectrum when using the default/optimal extraction aperture. The thin blue line represents the built-in effective area in the {\tt uvotpy} package with the first order anchor position of (1080, 1000) in detector pixels, and the thick orange line is derived with spectra listed in Table \ref{tbl:obs} with the mean anchor position of (988.4, 1080.2). Both shaded region represent the 1-$\sigma$ uncertainty.}
    \label{fig:ord2Arf}
\end{figure}

\begin{figure}[h]
    \centering
    \includegraphics[width=0.98\linewidth]{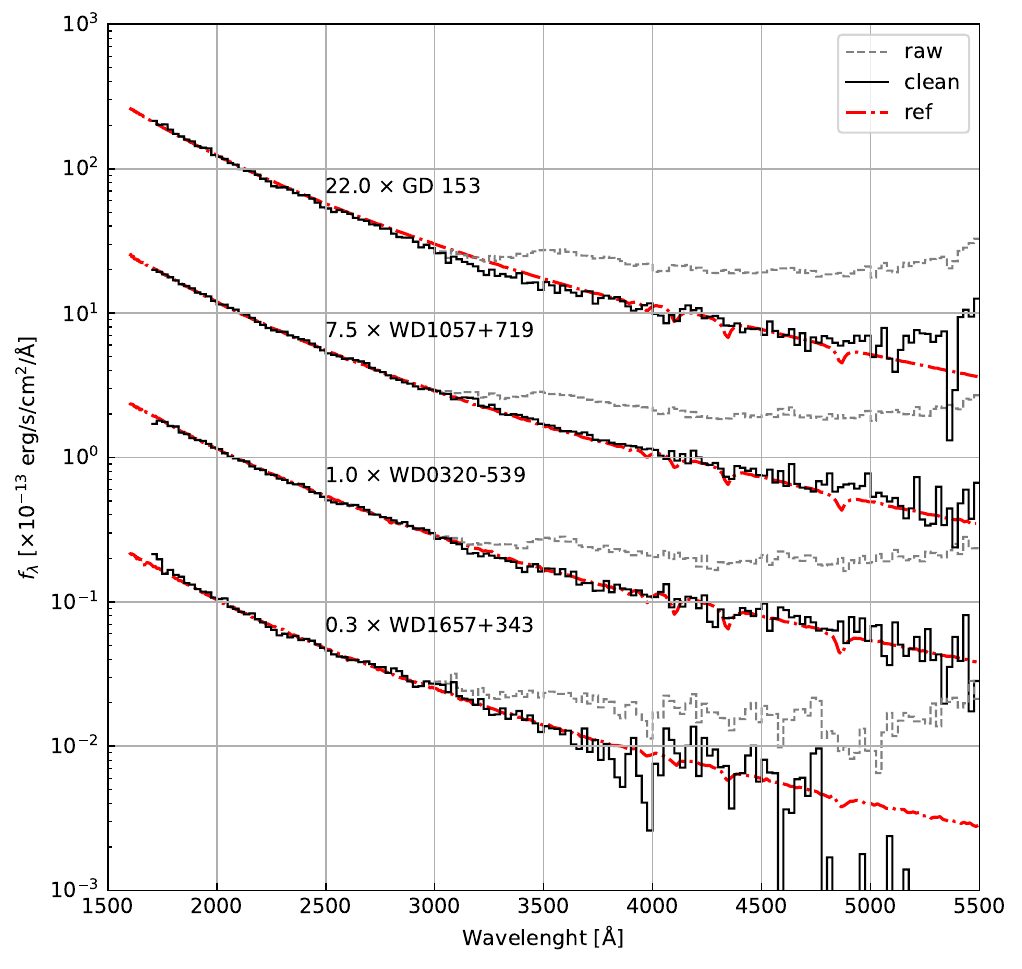}
    \caption{Cleaned first order spectra of the 4 white dwarfs used to calibrate the second order effective area. The black lines represent clean first order spectrum, and the gray dashed lines represent the raw first order spectrum extracted with \texttt{uvotpy}. Reference spectra are shown with red dot-dashed lines. The bin width is 25\,\AA. For GD 153, the spectrum suffers from the strong coincidence loss from $\sim3000\,\tAA$ to $\sim4000\,\tAA$, hence the intensity of the clean spectrum is a bit lower than the reference value. For WD1657+343, the signal-to-noise ratio is too low for $\lambda\gtrsim3800\,\tAA$, and the uncertainty is quite large, which is not plotted. Generally, the second order contamination is removed from the raw spectra, and the third order contamination for $\lambda\lesssim4500\,\tAA$ is negligible even for $\beta\sim-1.5$.}
    \label{fig:wdSpec}
\end{figure}

The reference spectra are taken from the CALSPEC \citep{2014PASP..126..711B,2020AJ....160...21B,2022stis.rept....7B,2022AJ....164...10B} and the method to derive the second order effective area is same as in \cite{2015MNRAS.449.2514K}:
\begin{itemize}
    \item [1)] Calculate the reference count rate of the first order with the reference spectrum.
    \item [2)] Subtract the reference count rate from the observed count rate to get the residual count rate.
    \item [3)] The residual count rate originates from the second order spectrum when the contribution of high orders is negligible, and the second order effective area can be calculated with Equation \ref{equ:cr2flux}. 
    %If ignoring the contribution of higher orders, the residual count rate originates from the second order and can be used to estimate the effective area with the reference spectrum, i.e., Equation \ref{equ:cr2flux}.
\end{itemize}
The calibrated second order effective area is shown in Figure \ref{fig:ord2Arf}, and the 1-$\sigma$ uncertainty for each wavelength bin is estimated from about 40 second order effective areas (observations contaminated with field stars at the given wavelength bin are excluded) calculated with individual observations, i.e., the 68.3\% confidence interval centered on the median. The cleaned first order spectra of the 4 white dwarfs selected to calibrate the second order effective area are shown in Figure \ref{fig:wdSpec}. The Clean Extraction successfully removes the second order contamination for $3000\,\tAA\lesssim\lambda\lesssim4000\,\tAA$.

\begin{figure}[h]
    \centering
    \includegraphics[width=0.49\linewidth]{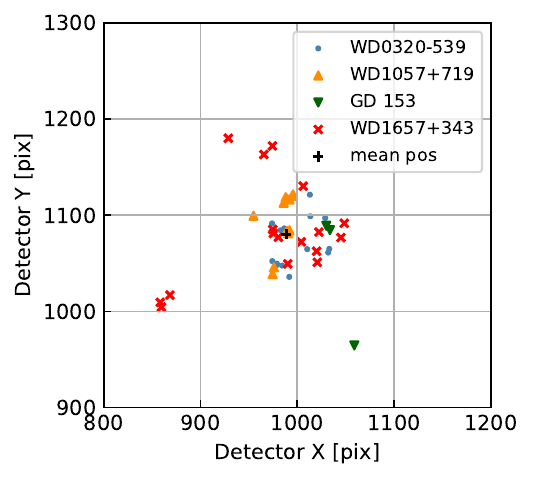}
    \caption{First order anchor positions of spectra listed in Table \ref{tbl:obs}. Different markers represent different targets, excepting the black cross, which represents the mean first anchor position of the all spectra used to calibrate the second order effective area.}
    \label{fig:dtpix}
\end{figure}

The anchor position influences the second order effective area, so it is necessary to define a region where the second order effective area is valid. The mean anchor position of the sources used for the calibration is (988.4, 1080.2), and the maximal distance of the sources from the mean anchor position is about 150 pixels. The anchor positions of spectra used to calibrate the second order effective area are shown in Figure \ref{fig:dtpix}. Hence, the second order effective area is valid in the circle region centered at (988.4, 1080.2) with a radius of $\sim150$ pixels.

\subsection{The systematic uncertainty}
\begin{figure}[h]
    \centering
    \includegraphics[width=0.49\linewidth]{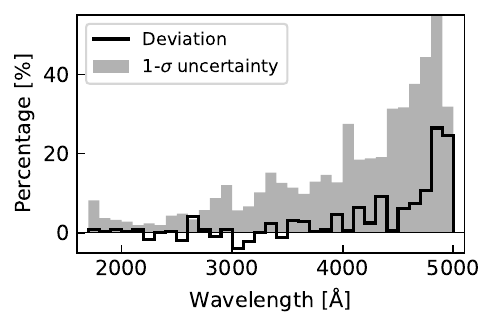}
    \caption{The deviation of cleaned first order spectra from reference spectra and the statistical 1-$\sigma$ uncertainty. The bin width of the wavelength is 100\,\AA, and the deviation is defined as the quotient of the residuals between cleaned spectra and reference spectra divided by reference spectra. For $\lambda<$2800\,\AA, the spectrum is free of the contamination. The median absolute deviation and the uncertainty are $\sim$0.8\,\% and $\sim$3.7\,\%, respectively. For 2800\,\AA$<\lambda<$4000\,\AA, the spectrum only suffers from the second order contamination and the median values are $\sim$1.3\,\% and $\sim$11.2\,\%, respectively. For $\lambda>$4000\,\AA, except the second order contamination, the spectrum also suffers from the third order contamination, hence the deviation roughly keeps becoming larger (especially for $\gtrsim4500\,\tAA$).}
    \label{fig:deviation}
\end{figure}

The total wavelength (from 1700 to 5000\,\tAA) is split into 33 bins with a width of 100\,\tAA~for two reasons: 1) the sample grid of the wavelength (depending on the anchor position) is different for each observation; 2) to get reliable statistics. Hence, the systematic uncertainty for each bin can be estimated with about 40 individual observations. The deviation between cleaned first order and reference spectra are shown in Figure \ref{fig:deviation}. For $\lambda<2800\,\tAA$, the median deviation and the median 1-$\sigma$ uncertainty (i.e., the 68.3\,\% quantile of the absolute deviations for each bin) are about 0.8\,\% and 3.7\,\%, respectively. For $2800\,\tAA<\lambda<4000\,\tAA$, the values are about 1.3\,\% and 11.2\,\%. For $\lambda>4000\,\tAA$, the deviation keeps going up and the 1-$\sigma$ uncertainty is quite large $\gtrsim30\,\%$ due to the third order contamination. Hence, we conclude for the wavelength range of $2800\,\tAA<\lambda<4000\,\tAA$, the systematic uncertainty of the Clean Extraction is about 11.2\,\%. %In addition, a more conservative value of 15\,\% can also be adopted, i.e., the maximal value of the 1-$\sigma$ uncertainty.

\section{Demonstration}\label{sec:demonstration}
During this discussion we simply adopt a power-law function to approximate the spectrum from $1700\,\tAA$ to $5000\,\tAA$:
\begin{equation}\label{equ:pl}
    f_\nu\propto\nu^{-\beta},
\end{equation}
where $\beta$ is the spectral index. Though the 4 white dwarfs are blue (all spectral indices are $\sim-1.5$), the third order contamination is negligible for $\lambda\lesssim4500\,\tAA$, because only a very small fraction of the third order trace overlaps with the extraction aperture.

\begin{figure}[h]
    \centering
    \includegraphics[width=0.49\linewidth]{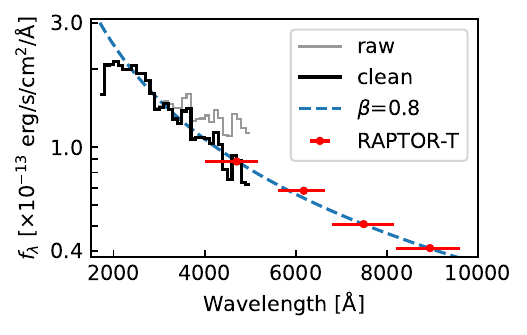}
    \caption{The cleaned first order spectrum of GRB 130427A. The gray and black lines are the raw first order spectrum extracted with \texttt{uvotpy} and the cleaned first order spectrum. The bin width is 100\,\AA. Red points represent photometries obtained by the RAPTOR-T telescope. The blue dashed line is the power-law function with $\beta=0.8$. The RAPTOR-T $g'$-band photometry matches the cleaned spectrum well, which implies the third order contamination is negligible for $4000\,\tAA\lesssim\lambda\lesssim5000\,\tAA$.}
    \label{fig:130427A}
\end{figure}

The RAPTOR (RAPid Telescopes for Optical Response) full sky monitoring telescopes \citep{2010SPIE.7737E..23W} captured the burst of GRB 130427A in the optical band and rapidly performed multi-band follow-up observations of GRB 130427A \citep{2014Sci...343...38V}. The Swift/UVOT observed the field of GRB 130427A from $\sim303$\,s to $\sim353\,$s after the BAT trigger with the UV grism in the nominal mode \citep{2014Sci...343...48M}. Hence, GRB 130427A is a good target to test the Clean Extraction. Figure \ref{fig:130427A} shows the comparison of the cleaned first order spectrum and the broad band SED simultaneously obtained by the RAPTOR-T. The cleaned UVOT spectrum of GRB 130427A is correlated with the transmission curve of the SDSS $g'$ band from $3630\,\tAA$ to $5830\,\tAA$. The spectral $g'$-band photometry is $11.94\pm0.22$ mag (AB), and the uncertainty should be at least 20\% (i.e., the 1-$\sigma$ uncertainty of the Clean Extraction for $\lambda>4000\,\tAA$). Because 1) we do not know the exact difference between SDSS $g'$ and RAPTOR $g'$, and 2) the correlated range is beyond the valid range of the UV grism spectrum, where the flux calibration could be inaccurate. From $\sim301\,$s to $\sim354\,$s after the BAT trigger, the RAPTOR-T $g'$-band photometry changes from $\sim11.93$ mag (AB) to $\sim12.14$ mag (AB), thus the RAPTOR-T $g'$-band photometry during the exposure of the UVOT spectrum is about $12.03\pm0.10$ mag (AB), which is consistent with the spectral $g'$-band photometry of $11.94\pm0.22$ mag (AB). While the spectral $g'$-band photometry derived with the uncleaned spectrum is about 11.48 mag (AB). Hence, we conclude that our method successfully removed the second order contamination. In addition, the third order contamination is actually negligible for $\beta\sim0.8$. Please note the anchor position of GRB 130427A is about 220 pixels away from the mean anchor position of the calibrated second order effective area, but the broad band SED shows that the Clean Extraction works well for GRB 130427A. If there is no reference spectrum or simultaneous photometry to check the validity of the cleaned spectrum, please be cautious of the cleaned spectrum with anchor positions falling outside the applicable region.

\begin{figure}[h]
    \centering
    \includegraphics[width=0.49\linewidth]{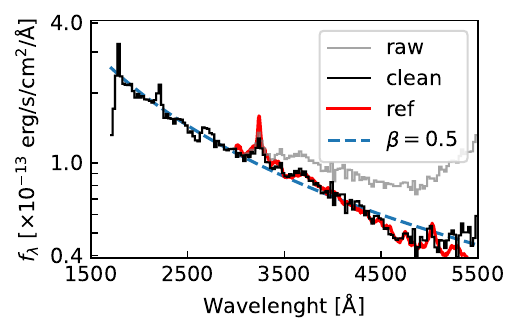}
    \caption{The cleaned first order spectrum of 3C273. The raw and cleaned spectra are shown with the gray and black lines respectively, and the bin width of the wavelength is 25\,\AA. The reference spectrum shown with the red line is taken by HST/STIS on Jan 31, 1999, and to match the intensity of UVOT UV grism spectrum, the reference spectrum is multiplied by a factor of 1.6. The blue dashed line is the power-law function with $\beta=0.5$. The higher order contamination begins to dominate at $\lambda\sim5100\,\text{\AA}$.}
    \label{fig:3C273}
\end{figure}

Swift/UVOT observed 3C 273 on Dec 13, 2005, however the reference spectrum was taken on Jan 31, 1999 (PI: John Hutchings, Proposal ID: 7568) with the space telescope imaging spectrograph on board the Hubble space telescope\footnote{The {\it HST} data used in this paper can be found in MAST: \dataset[10.17909/qd2p-1p70]{http://dx.doi.org/10.17909/qd2p-1p70}}. Table \ref{tbl:3C273} lists UVOT grism observations used in the paper. Though the luminosity of 3C 273 varies in a wide range, but the variability of its UV/optical spectral shape is less than $\sim10\,\%$ \citep[$3000\,\text{\AA}<\lambda<5000\,\tAA$, ][]{2008A&A...486..411S}, which is less than the systematic uncertainty of the Clean Extraction. Hence, 3C 273 is also a proper target to test the Clean Extraction. Figure \ref{fig:3C273} shows the comparison of the cleaned spectrum and the reference spectrum of 3C 273. The variability of 3C 273 in the UV/optical brightness is about a few tens percent within years \citep{2008A&A...486..411S}. Hence to account for the change in the brightness, the reference spectrum is multiplied by a factor of 1.6. The spectral index $\beta$ is $\sim0.5$, and the third order contamination is not important for $\lambda<5000\,\tAA$.

\subsection{The degree of the order 2 contamination}
\begin{figure}[h]
    \centering
    \includegraphics[width=0.49\linewidth]{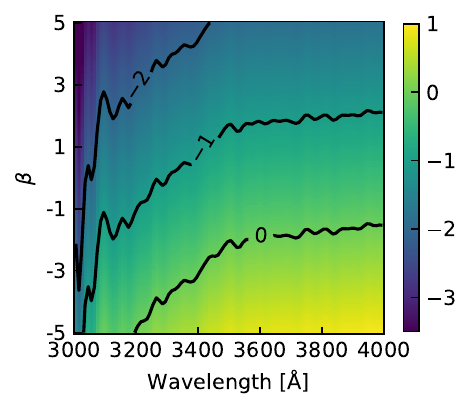}
    \caption{The logarithm of CR2/CR1 to base 10 for different spectral indices (y-axis) at a specified wavelength (x-axis). The black lines represent contours and the numbers in the black lines represent values of the contours. For $3600\,\tAA<\lambda<4000\,\tAA$, the second order count rate is greater than 10\,\% of the first order count rate when $\beta<2$.}
    \label{fig:O2dO1_ratio}
\end{figure}

The ratio of the second order count rate to the first order count rate, i.e., ${\rm CR_2}/{\rm CR_1}$ is a good indicator to quantify the degree of the second order contamination. Figure \ref{fig:O2dO1_ratio} shows how the logarithm of ${\rm CR_2}/{\rm CR_1}$ to base 10 evolves with the wavelength and the spectral index $\beta$. For $3600\,\tAA<\lambda<4000\,\tAA$, when $\beta\sim2$ (i.e., $f_\lambda\sim{\rm const}$), the second order count rate is about 10\,\% of the first order count rate, and when $\beta\sim-3$ (i.e., the approximation of the thermal emission at long wavelength), the second order count rate dominates. 

The degree of the second order contamination is listed in this Table \ref{tbl:look_up}, and colors of Swift/UVOT filters with different spectral indices are listed for the quick reference. For some grism observations, Swift/UVOT takes images before and after the spectroscopic exposures, which are called acquisition images. If the acquisition images are taken in different bands, one can derive the color of the target and find out the degree of the second order contamination for a power-law like SED in Table \ref{tbl:look_up}. Though the table is derived with the power-law SED, the color U-W2 or U-M2 works almost for arbitrary SEDs: the second order contamination occurs in the U-band wavelength: $\sim3000\,\tAA$ to $\sim4000\,\tAA$, and the contamination originates in W2 ($\sim1700\,\tAA$ to $\sim2200\,\tAA$) and M2 ($\sim2000\,\tAA$ to $\sim2500\,\tAA$) bands. Hence, the color U-W2 or U-M2 is the most direct indicator to trace the degree of the second order contamination.

For afterglows of gamma-ray bursts (GRBs), the intrinsic spectral shape can be described by the power-law function, and the typical value of the spectral index ranges from $\sim0.6$ to $\sim1.1$. For the prompt emission, the spectrum could be harder, hence in GRB/afterglow spectra, the second order contamination can not be ignored unless there is very strong UV/optical absorption, which presents further analysis of the spectrum with $\lambda\gtrsim3000\,\tAA$. However, the cleaned spectra of GRB 130427A and 3C273 shows that the third order contamination is negligible for typical spectral indices of GRBs.

\section{Conclusion}\label{sec:con}
Based on the output and calibration files of {\tt uvotpy} package, we proposed a method to remove the second order contamination from the Swift/UVOT UV nominal spectra. 4 white dwarfs are selected to calibrate the second order effective area and we found that the derived value is systematically smaller than the value in the {\tt uvotpy} package by about $2~{\rm cm}^2$. The typical 1-$\sigma$ systematic uncertainty of the cleaned spectrum in the contaminated region ($2800\,\tAA<\lambda<4000\,\tAA$) is $\sim11.2\%$. The red end of the cleaned spectrum is limited by the third order contamination. For blue sources, e.g., $\beta\sim-1.5$, the third order contamination becomes important at $\sim4500\,\tAA$, while for sources with $\beta\gtrsim0.5$, the third order contamination is negligible up to $5000\,\tAA$.

Currently, the Clean Extraction method is only valid for point sources, because the second order effective area defined in this article (please refer to the point 6 below) will be influenced by shapes of extended sources. To extract a cleaned UV nominal spectrum, make sure the anchor position of the source is located in the applicable region, i.e., a circle region centered at (988.4, 1080.2) in detector pixels with a radius of 150 pixels, and the extraction aperture should be the default/optimal aperture defined by the {\tt uvotpy} package. The reasons are summarized as follows:
\begin{itemize}
    \item [1)] This method uses the optimal/standard aperture (i.e., the trace of the 1st order spectrum) to extract the spectrum.
    \item [2)] As defined in \cite{2015MNRAS.449.2514K}, the optimal/standard aperture considers effects of the curvature of the first order at different anchor positions and the spatial point spread function (SPSF, the spread profile in the spatial direction at a specific wavelength) of the first order at different wavelengths.
    \item [3)] The half width of the optimal/standard aperture is $2.5\sigma$, where $\sigma$ is the standard deviation of the best fitted Gaussian function for the SPSF. Hence, when using the optimal/standard aperture to extract the spectrum, the first order effective area is always reliable.
    \item [4)] The spatial and dispersive offset between the second order and the first order depend on the anchor position\footnote{Please refer to \url{https://mssl.ucl.ac.uk/~npmk/Grism/order_layout.html} for more details}.
    \item [5)] When using the optimal/standard aperture to extract the spectrum, the ratio of the second order count falls in the optimal/standard aperture to the total second order rate (designated as R2) is different even at the same second order wavelength, because R2 is affected by a combination of the curvature and dispersion of the first and the second orders.
    \item [6)] The most accurate definition of the second order effective area in the Clean Extraction is the effective area that can be used to calculate the factor converting the second order rate within the optimal/standard aperture to the physical flux density, when using the optimal/standard aperture to extract the spectrum.
    \item [7)] Hence, it is necessary to define a specific region where the second order effective area does not deviate much from the one calibrated in the article, so that the Clean Extraction is applicable.
\end{itemize}

The work flow of the Clean Extraction is as follows:
\begin{itemize}
    \item [1)] Retrieve the keywords from the log in the header of the PHA file generated by the {\tt uvotpy} package.
    \item [2)] From the keywords, get the coefficients that convert pixel numbers to the wavelength for the first and the second order.
    \item [3)] From the keywords, get the associated effective area files and calculate the factor converting the count rate to flux density for the first and the second order.
    \item [4)] Resample the first spectrum to the wavelength range of the second order and estimate the count rate of the second order.
    \item [5)] Subtract the estimated count rate of the second order from the observed count rate, then multiply the residual count rate with the factor converting the count rate to flux density of the first order to get the cleaned spectrum.
\end{itemize}

The living source code is available on GitHub\footnote{\url{https://github.com/HaoZhou0810/cluvotpy}}, and the deposited version is available on Zenodo \citep{hao_zhou_2025_14603585}.

\newpage

\newpage
\begin{table}[ht!]
\centering
\caption{Observations used to calibrate the second order effective area for the UV nominal mode. The column ``OBS ID" is the unique observation id in the Swift archive and the column ``EXT" represents the extension id of the grism image file. The column ``DATE" is the start time of the observation.}\label{tbl:obs}
\begin{tabular}{ccccccc}
\hline
\hline
Source name & Spectral type & OBS ID & EXT & DATE & EXP & ${\rm ANK_1}$ \\
 & & & & (UTC) & (s) & (pixel) \\
\hline
WD0320-539 & DA1.5c  & 00054250001 & 1 & 2005-03-13T21:16:23 & 154.40 & (973.86, 1091.47) \\
 &  & 00054250001 & 2 & 2005-03-13T21:32:17 & 104.71 & (986.48, 1086.34) \\
 &  & 00054250001 & 3 & 2005-03-14T06:57:24 & 865.65 & (982.79, 1084.32) \\
 &  & 00054250001 & 4 & 2005-03-14T18:21:44 & 491.22 & (991.66, 1035.88) \\
 &  & 00054250001 & 5 & 2005-03-14T23:02:23 & 807.15 & (1010.23, 1064.69) \\
 &  & 00054250004 & 1 & 2005-03-31T00:41:23 & 590.83 & (1031.89, 1061.18) \\
 &  & 00054250004 & 2 & 2005-03-31T02:18:23 & 612.10 & (1033.01, 1065.00) \\
 &  & 00054250008 & 1 & 2005-05-12T18:18:46 & 607.63 & (979.02, 1048.32) \\
 &  & 00054250008 & 2 & 2005-05-12T19:55:46 & 548.60 & (974.14, 1052.28) \\
 &  & 00054250009 & 1 & 2005-05-12T21:32:47 & 548.19 & (978.88, 1049.64) \\
 &  & 00054250009 & 2 & 2005-05-12T23:08:47 & 539.52 & (984.33, 1047.61) \\
 &  & 00054250019 & 1 & 2005-10-23T16:45:24 & 1217.72 & (1028.79, 1097.16) \\
 &  & 00054250019 & 2 & 2005-10-23T18:22:24 & 1218.11 & (1012.95, 1121.34) \\
 &  & 00054250019 & 3 & 2005-10-23T20:02:24 & 981.95 & (1013.43, 1098.91) \\
\hline
WD1057+719 & DA1.2c & 00055200001 & 1 & 2005-03-14T20:07:24 & 1456.34 & (991.96, 1116.23) \\
 &  & 00055200001 & 2 & 2005-03-14T23:20:24 & 1042.88 & (990.50, 1081.14) \\
 &  & 00055200005 & 1 & 2005-03-13T23:12:52 & 1605.47 & (995.67, 1121.66) \\
 &  & 00055200010 & 1 & 2005-03-30T00:56:23 & 978.54 & (974.50, 1038.82) \\
 &  & 00055200010 & 2 & 2005-03-30T02:32:23 & 689.04 & (975.83, 1045.86) \\
 &  & 00055200016 & 1 & 2006-03-12T19:35:12 & 1701.00 & (991.96, 1084.33) \\
 &  & 00055200016 & 2 & 2006-03-12T21:11:37 & 1734.85 & (989.36, 1083.80) \\
 &  & 00055200020 & 1 & 2010-07-25T00:23:51 & 862.12 & (954.68, 1099.51) \\
 &  & 00055200032 & 1 & 2015-06-24T07:47:01 & 912.14 & (989.57, 1117.05) \\
 &  & 00055200032 & 2 & 2015-06-24T09:20:15 & 1059.79 & (985.86, 1112.88) \\
 &  & 00055200032 & 3 & 2015-06-24T14:07:15 & 1059.77 & (994.70, 1120.24) \\
 &  & 00055200032 & 4 & 2015-06-24T15:46:15 & 1059.79 & (987.91, 1118.76) \\
\hline
GD153 & DA1.2c & 00055500010 & 1 & 2005-04-12T01:35:44 & 195.96 & (1058.90, 964.55) \\
 &  & 00055500016 & 1 & 2005-11-09T01:39:24 & 1033.92 & (1029.93, 1089.00) \\
 &  & 00055500016 & 2 & 2005-11-09T03:16:23 & 1057.98 & (1033.65, 1084.45) \\
\hline
WD1657+343 & DA.9c & 00055900053 & 1 & 2008-05-11T11:43:37 & 495.83 & (868.35, 1016.94) \\
 &  & 00055900053 & 2 & 2008-05-11T13:20:37 & 466.29 & (858.37, 1009.38) \\
 &  & 00055900053 & 3 & 2008-05-11T14:56:37 & 495.82 & (859.58, 1004.29) \\
 &  & 00055900054 & 1 & 2008-05-13T05:06:55 & 612.95 & (1006.11, 1130.32) \\
 &  & 00055900056 & 1 & 2008-05-22T12:29:22 & 289.14 & (974.54, 1085.23) \\
 &  & 00055900056 & 2 & 2008-05-22T14:06:23 & 289.14 & (975.34, 1080.99) \\
 &  & 00055900057 & 2 & 2008-07-23T07:18:49 & 624.76 & (1048.49, 1091.63) \\
 &  & 00055900057 & 5 & 2008-07-23T08:55:49 & 624.76 & (965.41, 1163.10) \\
 &  & 00055900057 & 8 & 2008-07-23T10:31:48 & 624.76 & (974.27, 1172.05) \\
 &  & 00055900057 & 11 & 2008-07-23T12:08:48 & 624.76 & (928.68, 1180.06) \\
 &  & 00055900069 & 1 & 2012-06-14T18:42:58 & 315.87 & (1019.89, 1062.53) \\
 &  & 00055900069 & 2 & 2012-06-14T19:01:14 & 516.16 & (1004.10, 1072.27) \\
 &  & 00055900069 & 3 & 2012-06-14T22:03:07 & 969.23 & (990.08, 1049.40) \\
 &  & 00055900071 & 1 & 2012-07-31T08:40:21 & 1115.88 & (980.43, 1077.03) \\
 &  & 00055900071 & 2 & 2012-07-31T10:09:56 & 1460.37 & (1045.10, 1076.77) \\
 &  & 00055900071 & 3 & 2012-07-31T11:45:55 & 1461.35 & (1020.66, 1051.03) \\
 &  & 00055900071 & 4 & 2012-07-31T13:20:56 & 869.82 & (1022.32, 1082.47) \\
\hline
\hline
\end{tabular}
\end{table}

\newpage
\begin{table}[ht]
\centering
\caption{UVOT UV grism observations of 3C273. The observation ID is 00035017006. Definitions of columns are same as those in Table \ref{tbl:obs}.}\label{tbl:3C273}
\begin{tabular}{cccc}
\hline
\hline
EXT & DATE & EXP & ${\rm ANK_1}$ \\
& (UTC) & (s) & (pixel) \\
\hline
1 & 2005-12-13T00:24:23 & 1077.40 & (1047.14, 1060.55) \\
2 & 2005-12-13T02:06:24 & 730.45 & (979.85, 1112.11) \\
3 & 2005-12-13T03:42:23 & 729.66 & (982.48, 1106.82) \\
4 & 2005-12-13T05:18:24 & 773.89 & (998.42, 1094.04) \\
5 & 2005-12-13T06:55:24 & 730.95 & (984.12, 1111.15) \\
6 & 2005-12-13T16:37:23 & 1057.95 & (987.18, 1111.50) \\
7 & 2005-12-13T18:13:24 & 1191.67 & (994.64, 1105.37) \\
\hline
\hline
\end{tabular}
\end{table}

\begin{table}[h]
\centering
\caption{The degree of the second order contamination for power-law SEDs. For left to right, the first column is the spectral index and the next 4 columns are ratios of the second order count rate to the first order count rate in different wave bands. The left 6 columns are colors of Swift/UVOT filters in the AB system for the quick reference.}\label{tbl:look_up}
\begin{tabular}{c|cccc|cccccc}
\hline
\hline
$\beta$ & \multicolumn{4}{c|}{${\rm CR_2}/{\rm CR_1}$} & V-B & B-U & U-W1 & W1-M2 & M2-W2 & W2-WH \\
 & (3000-3200\,\AA) & (3200-3400\,\AA) & (3400-3600\,\AA) & (3600-4000\,\AA) & (AB) & (AB) & (AB) & (AB) & (AB) & (AB) \\
\hline
-5.0 & 0.56 & 2.06 & 5.36 & 7.72 & 1.25 & 1.23 & 1.82 & 0.58 & 0.69 & -1.00 \\
-4.5 & 0.41 & 1.53 & 3.94 & 5.65 & 1.12 & 1.11 & 1.62 & 0.54 & 0.62 & -0.96 \\ 
-4.0 & 0.31 & 1.13 & 2.90 & 4.13 & 1.00 & 0.99 & 1.42 & 0.50 & 0.54 & -0.91 \\ 
-3.5 & 0.23 & 0.83 & 2.14 & 3.02 & 0.87 & 0.86 & 1.23 & 0.45 & 0.47 & -0.86 \\ 
-3.0 & 0.17 & 0.62 & 1.57 & 2.21 & 0.74 & 0.74 & 1.04 & 0.40 & 0.40 & -0.79 \\ 
-2.5 & 0.13 & 0.46 & 1.16 & 1.62 & 0.62 & 0.62 & 0.85 & 0.34 & 0.33 & -0.71 \\
-2.0 & 0.09 & 0.34 & 0.85 & 1.18 & 0.49 & 0.49 & 0.67 & 0.28 & 0.26 & -0.61 \\
-1.5 & 0.07 & 0.25 & 0.63 & 0.86 & 0.37 & 0.37 & 0.50 & 0.21 & 0.19 & -0.50 \\
-1.0 & 0.05 & 0.18 & 0.46 & 0.63 & 0.24 & 0.25 & 0.33 & 0.15 & 0.12 & -0.36 \\
-0.5 & 0.04 & 0.14 & 0.34 & 0.46 & 0.12 & 0.12 & 0.16 & 0.08 & 0.06 & -0.19 \\
0.0 & 0.03 & 0.10 & 0.25 & 0.34 & 0.00 & 0.00 & 0.00 & 0.00 & 0.00 & 0.00 \\
0.5 & 0.02 & 0.07 & 0.18 & 0.25 & -0.12 & -0.12 & -0.16 & -0.08 & -0.06 & 0.22 \\
1.0 & 0.02 & 0.06 & 0.14 & 0.18 & -0.24 & -0.25 & -0.31 & -0.16 & -0.11 & 0.47 \\
1.5 & 0.01 & 0.04 & 0.10 & 0.13 & -0.36 & -0.37 & -0.46 & -0.25 & -0.17 & 0.76 \\
2.0 & 0.01 & 0.03 & 0.07 & 0.10 & -0.48 & -0.49 & -0.61 & -0.34 & -0.22 & 1.07 \\
2.5 & 0.01 & 0.02 & 0.05 & 0.07 & -0.60 & -0.61 & -0.75 & -0.43 & -0.26 & 1.40 \\
3.0 & 0.00 & 0.02 & 0.04 & 0.05 & -0.72 & -0.73 & -0.88 & -0.53 & -0.30 & 1.77 \\
3.5 & 0.00 & 0.01 & 0.03 & 0.04 & -0.83 & -0.86 & -1.02 & -0.63 & -0.34 & 2.15 \\
4.0 & 0.00 & 0.01 & 0.02 & 0.03 & -0.95 & -0.98 & -1.14 & -0.73 & -0.37 & 2.55 \\
4.5 & 0.00 & 0.01 & 0.02 & 0.02 & -1.07 & -1.10 & -1.26 & -0.85 & -0.40 & 2.97 \\
5.0 & 0.00 & 0.00 & 0.01 & 0.01 & -1.18 & -1.22 & -1.38 & -0.96 & -0.42 & 3.40 \\
\hline
\hline
\end{tabular}
\end{table}

\begin{acknowledgments}
We acknowledge the use of the public data from Swift archive, Mikulski Archive for Space Telescopes and CALSPEC. This work is supported by the Natural Science Foundation of China (NSFC) under grant Nos. 12225305, 12321003 and 12473049, the Strategic Priority Research Program of the Chinese Academy of Sciences, grant No. XDB0550400, the China Manned Space Project (No. CMS-CSST-2021-A12) and the Italian Space Agency, contract ASI/INAF n. I/004/11/6.
\end{acknowledgments}


\vspace{5mm}
\facilities{Swift(UVOT), HST(STIS)}


\software{astropy \citep{astropy:2013,astropy:2018,astropy:2022}, uvotpy \citep{2014ascl.soft10004K}, HEASoft \citep{2014ascl.soft08004N}}



%\appendix


\bibliography{sample631}{}

\begin{thebibliography}{}
\expandafter\ifx\csname natexlab\endcsname\relax\def\natexlab#1{#1}\fi
\providecommand{\url}[1]{\href{#1}{#1}}
\providecommand{\dodoi}[1]{doi:~\href{http://doi.org/#1}{\nolinkurl{#1}}}
\providecommand{\doeprint}[1]{\href{http://ascl.net/#1}{\nolinkurl{http://ascl.net/#1}}}
\providecommand{\doarXiv}[1]{\href{https://arxiv.org/abs/#1}{\nolinkurl{https://arxiv.org/abs/#1}}}

\bibitem[{{Astropy Collaboration} {et~al.}(2013){Astropy Collaboration},
  {Robitaille}, {Tollerud}, {Greenfield}, {Droettboom}, {Bray}, {Aldcroft},
  {Davis}, {Ginsburg}, {Price-Whelan}, {Kerzendorf}, {Conley}, {Crighton},
  {Barbary}, {Muna}, {Ferguson}, {Grollier}, {Parikh}, {Nair}, {Unther},
  {Deil}, {Woillez}, {Conseil}, {Kramer}, {Turner}, {Singer}, {Fox}, {Weaver},
  {Zabalza}, {Edwards}, {Azalee Bostroem}, {Burke}, {Casey}, {Crawford},
  {Dencheva}, {Ely}, {Jenness}, {Labrie}, {Lim}, {Pierfederici}, {Pontzen},
  {Ptak}, {Refsdal}, {Servillat}, \& {Streicher}}]{astropy:2013}
{Astropy Collaboration}, {Robitaille}, T.~P., {Tollerud}, E.~J., {et~al.} 2013,
  \aap, 558, A33, \dodoi{10.1051/0004-6361/201322068}

\bibitem[{{Astropy Collaboration} {et~al.}(2018){Astropy Collaboration},
  {Price-Whelan}, {Sip{\H{o}}cz}, {G{\"u}nther}, {Lim}, {Crawford}, {Conseil},
  {Shupe}, {Craig}, {Dencheva}, {Ginsburg}, {Vand erPlas}, {Bradley},
  {P{\'e}rez-Su{\'a}rez}, {de Val-Borro}, {Aldcroft}, {Cruz}, {Robitaille},
  {Tollerud}, {Ardelean}, {Babej}, {Bach}, {Bachetti}, {Bakanov}, {Bamford},
  {Barentsen}, {Barmby}, {Baumbach}, {Berry}, {Biscani}, {Boquien}, {Bostroem},
  {Bouma}, {Brammer}, {Bray}, {Breytenbach}, {Buddelmeijer}, {Burke},
  {Calderone}, {Cano Rodr{\'\i}guez}, {Cara}, {Cardoso}, {Cheedella}, {Copin},
  {Corrales}, {Crichton}, {D'Avella}, {Deil}, {Depagne}, {Dietrich}, {Donath},
  {Droettboom}, {Earl}, {Erben}, {Fabbro}, {Ferreira}, {Finethy}, {Fox},
  {Garrison}, {Gibbons}, {Goldstein}, {Gommers}, {Greco}, {Greenfield},
  {Groener}, {Grollier}, {Hagen}, {Hirst}, {Homeier}, {Horton}, {Hosseinzadeh},
  {Hu}, {Hunkeler}, {Ivezi{\'c}}, {Jain}, {Jenness}, {Kanarek}, {Kendrew},
  {Kern}, {Kerzendorf}, {Khvalko}, {King}, {Kirkby}, {Kulkarni}, {Kumar},
  {Lee}, {Lenz}, {Littlefair}, {Ma}, {Macleod}, {Mastropietro}, {McCully},
  {Montagnac}, {Morris}, {Mueller}, {Mumford}, {Muna}, {Murphy}, {Nelson},
  {Nguyen}, {Ninan}, {N{\"o}the}, {Ogaz}, {Oh}, {Parejko}, {Parley}, {Pascual},
  {Patil}, {Patil}, {Plunkett}, {Prochaska}, {Rastogi}, {Reddy Janga},
  {Sabater}, {Sakurikar}, {Seifert}, {Sherbert}, {Sherwood-Taylor}, {Shih},
  {Sick}, {Silbiger}, {Singanamalla}, {Singer}, {Sladen}, {Sooley},
  {Sornarajah}, {Streicher}, {Teuben}, {Thomas}, {Tremblay}, {Turner},
  {Terr{\'o}n}, {van Kerkwijk}, {de la Vega}, {Watkins}, {Weaver}, {Whitmore},
  {Woillez}, {Zabalza}, \& {Astropy Contributors}}]{astropy:2018}
{Astropy Collaboration}, {Price-Whelan}, A.~M., {Sip{\H{o}}cz}, B.~M., {et~al.}
  2018, \aj, 156, 123, \dodoi{10.3847/1538-3881/aabc4f}

\bibitem[{{Astropy Collaboration} {et~al.}(2022){Astropy Collaboration},
  {Price-Whelan}, {Lim}, {Earl}, {Starkman}, {Bradley}, {Shupe}, {Patil},
  {Corrales}, {Brasseur}, {N{"o}the}, {Donath}, {Tollerud}, {Morris},
  {Ginsburg}, {Vaher}, {Weaver}, {Tocknell}, {Jamieson}, {van Kerkwijk},
  {Robitaille}, {Merry}, {Bachetti}, {G{"u}nther}, {Aldcroft},
  {Alvarado-Montes}, {Archibald}, {B{'o}di}, {Bapat}, {Barentsen}, {Baz{'a}n},
  {Biswas}, {Boquien}, {Burke}, {Cara}, {Cara}, {Conroy}, {Conseil}, {Craig},
  {Cross}, {Cruz}, {D'Eugenio}, {Dencheva}, {Devillepoix}, {Dietrich},
  {Eigenbrot}, {Erben}, {Ferreira}, {Foreman-Mackey}, {Fox}, {Freij}, {Garg},
  {Geda}, {Glattly}, {Gondhalekar}, {Gordon}, {Grant}, {Greenfield}, {Groener},
  {Guest}, {Gurovich}, {Handberg}, {Hart}, {Hatfield-Dodds}, {Homeier},
  {Hosseinzadeh}, {Jenness}, {Jones}, {Joseph}, {Kalmbach}, {Karamehmetoglu},
  {Ka{l}uszy{'n}ski}, {Kelley}, {Kern}, {Kerzendorf}, {Koch}, {Kulumani},
  {Lee}, {Ly}, {Ma}, {MacBride}, {Maljaars}, {Muna}, {Murphy}, {Norman},
  {O'Steen}, {Oman}, {Pacifici}, {Pascual}, {Pascual-Granado}, {Patil},
  {Perren}, {Pickering}, {Rastogi}, {Roulston}, {Ryan}, {Rykoff}, {Sabater},
  {Sakurikar}, {Salgado}, {Sanghi}, {Saunders}, {Savchenko}, {Schwardt},
  {Seifert-Eckert}, {Shih}, {Jain}, {Shukla}, {Sick}, {Simpson},
  {Singanamalla}, {Singer}, {Singhal}, {Sinha}, {Sip{H{o}}cz}, {Spitler},
  {Stansby}, {Streicher}, {{{S}}umak}, {Swinbank}, {Taranu}, {Tewary},
  {Tremblay}, {Val-Borro}, {Van Kooten}, {Vasovi{'c}}, {Verma}, {de Miranda
  Cardoso}, {Williams}, {Wilson}, {Winkel}, {Wood-Vasey}, {Xue}, {Yoachim},
  {Zhang}, {Zonca}, \& {Astropy Project Contributors}}]{astropy:2022}
{Astropy Collaboration}, {Price-Whelan}, A.~M., {Lim}, P.~L., {et~al.} 2022,
  \apj, 935, 167, \dodoi{10.3847/1538-4357/ac7c74}

\bibitem[{{Barthelmy} {et~al.}(2005){Barthelmy}, {Barbier}, {Cummings},
  {Fenimore}, {Gehrels}, {Hullinger}, {Krimm}, {Markwardt}, {Palmer},
  {Parsons}, {Sato}, {Suzuki}, {Takahashi}, {Tashiro}, \&
  {Tueller}}]{2005SSRv..120..143B}
{Barthelmy}, S.~D., {Barbier}, L.~M., {Cummings}, J.~R., {et~al.} 2005, \ssr,
  120, 143, \dodoi{10.1007/s11214-005-5096-3}

\bibitem[{{Bohlin} {et~al.}(2014){Bohlin}, {Gordon}, \&
  {Tremblay}}]{2014PASP..126..711B}
{Bohlin}, R.~C., {Gordon}, K.~D., \& {Tremblay}, P.~E. 2014, \pasp, 126, 711,
  \dodoi{10.1086/677655}

\bibitem[{{Bohlin} {et~al.}(2020){Bohlin}, {Hubeny}, \&
  {Rauch}}]{2020AJ....160...21B}
{Bohlin}, R.~C., {Hubeny}, I., \& {Rauch}, T. 2020, \aj, 160, 21,
  \dodoi{10.3847/1538-3881/ab94b4}

\bibitem[{{Bohlin} {et~al.}(2022){Bohlin}, {Krick}, {Gordon}, \&
  {Hubeny}}]{2022AJ....164...10B}
{Bohlin}, R.~C., {Krick}, J.~E., {Gordon}, K.~D., \& {Hubeny}, I. 2022, \aj,
  164, 10, \dodoi{10.3847/1538-3881/ac6fe1}

\bibitem[{{Bohlin} \& {Lockwood}(2022)}]{2022stis.rept....7B}
{Bohlin}, R.~C., \& {Lockwood}, S. 2022, {Update of the STIS CTE Correction
  Formula for Stellar Spectra}, Instrument Science Report STIS 2022-7, 11 pages

\bibitem[{{Bostroem} {et~al.}(2019){Bostroem}, {Valenti}, {Horesh}, {Morozova},
  {Kuin}, {Wyatt}, {Jerkstrand}, {Sand}, {Lundquist}, {Smith}, {Sullivan},
  {Hosseinzadeh}, {Arcavi}, {Callis}, {Cartier}, {Gal-Yam}, {Galbany},
  {Guti{\'e}rrez}, {Howell}, {Inserra}, {Kankare}, {L{\'o}pez}, {McCully},
  {Pignata}, {Piro}, {Rodr{\'\i}guez}, {Smartt}, {Smith}, {Yaron}, \&
  {Young}}]{2019MNRAS.485.5120B}
{Bostroem}, K.~A., {Valenti}, S., {Horesh}, A., {et~al.} 2019, \mnras, 485,
  5120, \dodoi{10.1093/mnras/stz570}

\bibitem[{{Breeveld} {et~al.}(2011){Breeveld}, {Landsman}, {Holland}, {Roming},
  {Kuin}, \& {Page}}]{2011AIPC.1358..373B}
{Breeveld}, A.~A., {Landsman}, W., {Holland}, S.~T., {et~al.} 2011, in American
  Institute of Physics Conference Series, Vol. 1358, Gamma Ray Bursts 2010, ed.
  J.~E. {McEnery}, J.~L. {Racusin}, \& N.~{Gehrels} (AIP), 373--376,
  \dodoi{10.1063/1.3621807}

\bibitem[{{Brown} {et~al.}(2015){Brown}, {Smitka}, {Wang}, {Breeveld}, {de
  Pasquale}, {Hartmann}, {Krisciunas}, {Kuin}, {Milne}, {Page}, \&
  {Siegel}}]{2015ApJ...805...74B}
{Brown}, P.~J., {Smitka}, M.~T., {Wang}, L., {et~al.} 2015, \apj, 805, 74,
  \dodoi{10.1088/0004-637X/805/1/74}

\bibitem[{{Burrows} {et~al.}(2005){Burrows}, {Hill}, {Nousek}, {Kennea},
  {Wells}, {Osborne}, {Abbey}, {Beardmore}, {Mukerjee}, {Short}, {Chincarini},
  {Campana}, {Citterio}, {Moretti}, {Pagani}, {Tagliaferri}, {Giommi},
  {Capalbi}, {Tamburelli}, {Angelini}, {Cusumano}, {Br{\"a}uninger}, {Burkert},
  \& {Hartner}}]{2005SSRv..120..165B}
{Burrows}, D.~N., {Hill}, J.~E., {Nousek}, J.~A., {et~al.} 2005, \ssr, 120,
  165, \dodoi{10.1007/s11214-005-5097-2}

\bibitem[{{Gehrels} {et~al.}(2004){Gehrels}, {Chincarini}, {Giommi}, {Mason},
  {Nousek}, {Wells}, {White}, {Barthelmy}, {Burrows}, {Cominsky}, {Hurley},
  {Marshall}, {M{\'e}sz{\'a}ros}, {Roming}, {Angelini}, {Barbier}, {Belloni},
  {Campana}, {Caraveo}, {Chester}, {Citterio}, {Cline}, {Cropper}, {Cummings},
  {Dean}, {Feigelson}, {Fenimore}, {Frail}, {Fruchter}, {Garmire}, {Gendreau},
  {Ghisellini}, {Greiner}, {Hill}, {Hunsberger}, {Krimm}, {Kulkarni}, {Kumar},
  {Lebrun}, {Lloyd-Ronning}, {Markwardt}, {Mattson}, {Mushotzky}, {Norris},
  {Osborne}, {Paczynski}, {Palmer}, {Park}, {Parsons}, {Paul}, {Rees},
  {Reynolds}, {Rhoads}, {Sasseen}, {Schaefer}, {Short}, {Smale}, {Smith},
  {Stella}, {Tagliaferri}, {Takahashi}, {Tashiro}, {Townsley}, {Tueller},
  {Turner}, {Vietri}, {Voges}, {Ward}, {Willingale}, {Zerbi}, \&
  {Zhang}}]{2004ApJ...611.1005G}
{Gehrels}, N., {Chincarini}, G., {Giommi}, P., {et~al.} 2004, \apj, 611, 1005,
  \dodoi{10.1086/422091}

\bibitem[{{Guiriec} {et~al.}(2016){Guiriec}, {Kouveliotou}, {Hartmann},
  {Granot}, {Asano}, {M{\'e}sz{\'a}ros}, {Gill}, {Gehrels}, \&
  {McEnery}}]{2016ApJ...831L...8G}
{Guiriec}, S., {Kouveliotou}, C., {Hartmann}, D.~H., {et~al.} 2016, \apjl, 831,
  L8, \dodoi{10.3847/2041-8205/831/1/L8}

\bibitem[{{Ho} {et~al.}(2019){Ho}, {Goldstein}, {Schulze}, {Khatami}, {Perley},
  {Ergon}, {Gal-Yam}, {Corsi}, {Andreoni}, {Barbarino}, {Bellm},
  {Blagorodnova}, {Bright}, {Burns}, {Cenko}, {Cunningham}, {De}, {Dekany},
  {Dugas}, {Fender}, {Fransson}, {Fremling}, {Goldstein}, {Graham}, {Hale},
  {Horesh}, {Hung}, {Kasliwal}, {Kuin}, {Kulkarni}, {Kupfer}, {Lunnan},
  {Masci}, {Ngeow}, {Nugent}, {Ofek}, {Patterson}, {Petitpas}, {Rusholme},
  {Sai}, {Sfaradi}, {Shupe}, {Sollerman}, {Soumagnac}, {Tachibana}, {Taddia},
  {Walters}, {Wang}, {Yao}, \& {Zhang}}]{2019ApJ...887..169H}
{Ho}, A. Y.~Q., {Goldstein}, D.~A., {Schulze}, S., {et~al.} 2019, \apj, 887,
  169, \dodoi{10.3847/1538-4357/ab55ec}

\bibitem[{{Jin} {et~al.}(2023){Jin}, {Zhou}, {Wang}, {Geng}, {Covino}, {Wu},
  {Li}, {Fan}, {Wei}, \& {Wei}}]{2023NatAs...7.1108J}
{Jin}, Z.-P., {Zhou}, H., {Wang}, Y., {et~al.} 2023, Nature Astronomy, 7, 1108,
  \dodoi{10.1038/s41550-023-02005-w}

\bibitem[{{Kuin} {et~al.}(2009){Kuin}, {Landsman}, {Page}, {Schady}, {Still},
  {Breeveld}, {de Pasquale}, {Roming}, {Brown}, {Carter}, {James}, {Curran},
  {Cucchiara}, {Gronwall}, {Holland}, {Hoversten}, {Hunsberger}, {Kennedy},
  {Koch}, {Lamoureux}, {Marshall}, {Oates}, {Parsons}, {Palmer}, \&
  {Smith}}]{2009MNRAS.395L..21K}
{Kuin}, N.~P.~M., {Landsman}, W., {Page}, M.~J., {et~al.} 2009, \mnras, 395,
  L21, \dodoi{10.1111/j.1745-3933.2009.00632.x}

\bibitem[{{Kuin} {et~al.}(2015){Kuin}, {Landsman}, {Breeveld}, {Page},
  {Lamoureux}, {James}, {Mehdipour}, {Still}, {Yershov}, {Brown}, {Carter},
  {Mason}, {Kennedy}, {Marshall}, {Roming}, {Siegel}, {Oates}, {Smith}, \& {De
  Pasquale}}]{2015MNRAS.449.2514K}
{Kuin}, N.~P.~M., {Landsman}, W., {Breeveld}, A.~A., {et~al.} 2015, \mnras,
  449, 2514, \dodoi{10.1093/mnras/stv408}

\bibitem[{{Kuin} {et~al.}(2019){Kuin}, {Wu}, {Oates}, {Lien}, {Emery},
  {Kennea}, {de Pasquale}, {Han}, {Brown}, {Tohuvavohu}, {Breeveld}, {Burrows},
  {Cenko}, {Campana}, {Levan}, {Markwardt}, {Osborne}, {Page}, {Page},
  {Sbarufatti}, {Siegel}, \& {Troja}}]{2019MNRAS.487.2505K}
{Kuin}, N. P.~M., {Wu}, K., {Oates}, S., {et~al.} 2019, \mnras, 487, 2505,
  \dodoi{10.1093/mnras/stz053}

\bibitem[{{Kuin} {et~al.}(2020){Kuin}, {Page}, {Mr{\'o}z}, {Darnley}, {Shore},
  {Osborne}, {Walter}, {Di Mille}, {Morrell}, {Munari}, {Bohlsen}, {Evans},
  {Gehrz}, {Starrfield}, {Henze}, {Williams}, {Schwarz}, {Udalski},
  {Szyma{\'n}ski}, {Poleski}, {Soszy{\'n}ski}, {Ribeiro}, {Angeloni},
  {Breeveld}, {Beardmore}, \& {Skowron}}]{2020MNRAS.491..655K}
{Kuin}, N.~P.~M., {Page}, K.~L., {Mr{\'o}z}, P., {et~al.} 2020, \mnras, 491,
  655, \dodoi{10.1093/mnras/stz2960}

\bibitem[{{Kuin}(2014)}]{2014ascl.soft10004K}
{Kuin}, P. 2014, {UVOTPY: Swift UVOT grism data reduction}, Astrophysics Source
  Code Library, record ascl:1410.004

\bibitem[{{Maselli} {et~al.}(2014){Maselli}, {Melandri}, {Nava}, {Mundell},
  {Kawai}, {Campana}, {Covino}, {Cummings}, {Cusumano}, {Evans}, {Ghirlanda},
  {Ghisellini}, {Guidorzi}, {Kobayashi}, {Kuin}, {La Parola}, {Mangano},
  {Oates}, {Sakamoto}, {Serino}, {Virgili}, {Zhang}, {Barthelmy}, {Beardmore},
  {Bernardini}, {Bersier}, {Burrows}, {Calderone}, {Capalbi}, {Chiang},
  {D'Avanzo}, {D'Elia}, {De Pasquale}, {Fugazza}, {Gehrels}, {Gomboc},
  {Harrison}, {Hanayama}, {Japelj}, {Kennea}, {Kopac}, {Kouveliotou}, {Kuroda},
  {Levan}, {Malesani}, {Marshall}, {Nousek}, {O'Brien}, {Osborne}, {Pagani},
  {Page}, {Page}, {Perri}, {Pritchard}, {Romano}, {Saito}, {Sbarufatti},
  {Salvaterra}, {Steele}, {Tanvir}, {Vianello}, {Weigand}, {Wiersema}, {Yatsu},
  {Yoshii}, \& {Tagliaferri}}]{2014Sci...343...48M}
{Maselli}, A., {Melandri}, A., {Nava}, L., {et~al.} 2014, Science, 343, 48,
  \dodoi{10.1126/science.1242279}

\bibitem[{{Mason} {et~al.}(2004){Mason}, {Breeveld}, {Hunsberger}, {James},
  {Kennedy}, {Roming}, \& {Stock}}]{2004SPIE.5165..277M}
{Mason}, K.~O., {Breeveld}, A., {Hunsberger}, S.~D., {et~al.} 2004, in Society
  of Photo-Optical Instrumentation Engineers (SPIE) Conference Series, Vol.
  5165, X-Ray and Gamma-Ray Instrumentation for Astronomy XIII, ed. K.~A.
  {Flanagan} \& O.~H.~W. {Siegmund}, 277--286, \dodoi{10.1117/12.503713}

\bibitem[{{Nasa High Energy Astrophysics Science Archive Research Center
  (Heasarc)}(2014)}]{2014ascl.soft08004N}
{Nasa High Energy Astrophysics Science Archive Research Center (Heasarc)}.
  2014, {HEAsoft: Unified Release of FTOOLS and XANADU}, Astrophysics Source
  Code Library, record ascl:1408.004

\bibitem[{{Page} {et~al.}(2013){Page}, {Kuin}, {Breeveld}, {Hancock},
  {Holland}, {Marshall}, {Oates}, {Roming}, {Siegel}, {Smith}, {Carter}, {De
  Pasquale}, {Symeonidis}, {Yershov}, \& {Beardmore}}]{2013MNRAS.436.1684P}
{Page}, M.~J., {Kuin}, N.~P.~M., {Breeveld}, A.~A., {et~al.} 2013, \mnras, 436,
  1684, \dodoi{10.1093/mnras/stt1689}

\bibitem[{{Pan} {et~al.}(2018){Pan}, {Foley}, {Filippenko}, \&
  {Kuin}}]{2018MNRAS.479..517P}
{Pan}, Y.~C., {Foley}, R.~J., {Filippenko}, A.~V., \& {Kuin}, N.~P.~M. 2018,
  \mnras, 479, 517, \dodoi{10.1093/mnras/sty1420}

\bibitem[{{Pan} {et~al.}(2020){Pan}, {Foley}, {Jones}, {Filippenko}, \&
  {Kuin}}]{2020MNRAS.491.5897P}
{Pan}, Y.~C., {Foley}, R.~J., {Jones}, D.~O., {Filippenko}, A.~V., \& {Kuin},
  N.~P.~M. 2020, \mnras, 491, 5897, \dodoi{10.1093/mnras/stz3391}

\bibitem[{{Pirzkal} {et~al.}(2017){Pirzkal}, {Hilbert}, \&
  {Rothberg}}]{2017wfc..rept...20P}
{Pirzkal}, N., {Hilbert}, B., \& {Rothberg}, B. 2017, {Trace and Wavelength
  Calibrations of the UVIS G280 +1/-1 Grism Orders}, Instrument Science Report
  WFC3 2017-20, 15 pages

\bibitem[{{Poole} {et~al.}(2008){Poole}, {Breeveld}, {Page}, {Landsman},
  {Holland}, {Roming}, {Kuin}, {Brown}, {Gronwall}, {Hunsberger}, {Koch},
  {Mason}, {Schady}, {vanden Berk}, {Blustin}, {Boyd}, {Broos}, {Carter},
  {Chester}, {Cucchiara}, {Hancock}, {Huckle}, {Immler}, {Ivanushkina},
  {Kennedy}, {Marshall}, {Morgan}, {Pandey}, {de Pasquale}, {Smith}, \&
  {Still}}]{2008MNRAS.383..627P}
{Poole}, T.~S., {Breeveld}, A.~A., {Page}, M.~J., {et~al.} 2008, \mnras, 383,
  627, \dodoi{10.1111/j.1365-2966.2007.12563.x}

\bibitem[{{Roming} {et~al.}(2005){Roming}, {Kennedy}, {Mason}, {Nousek}, {Ahr},
  {Bingham}, {Broos}, {Carter}, {Hancock}, {Huckle}, {Hunsberger}, {Kawakami},
  {Killough}, {Koch}, {McLelland}, {Smith}, {Smith}, {Soto}, {Boyd},
  {Breeveld}, {Holland}, {Ivanushkina}, {Pryzby}, {Still}, \&
  {Stock}}]{2005SSRv..120...95R}
{Roming}, P. W.~A., {Kennedy}, T.~E., {Mason}, K.~O., {et~al.} 2005, \ssr, 120,
  95, \dodoi{10.1007/s11214-005-5095-4}

\bibitem[{{Smitka} {et~al.}(2015){Smitka}, {Brown}, {Suntzeff}, {Zhang},
  {Zhai}, {Wang}, {Mo}, \& {Zhang}}]{2015ApJ...813...30S}
{Smitka}, M.~T., {Brown}, P.~J., {Suntzeff}, N.~B., {et~al.} 2015, \apj, 813,
  30, \dodoi{10.1088/0004-637X/813/1/30}

\bibitem[{{Soldi} {et~al.}(2008){Soldi}, {T{\"u}rler}, {Paltani}, {Aller},
  {Aller}, {Burki}, {Chernyakova}, {L{\"a}hteenm{\"a}ki}, {McHardy}, {Robson},
  {Staubert}, {Tornikoski}, {Walter}, \& {Courvoisier}}]{2008A&A...486..411S}
{Soldi}, S., {T{\"u}rler}, M., {Paltani}, S., {et~al.} 2008, \aap, 486, 411,
  \dodoi{10.1051/0004-6361:200809947}

\bibitem[{{Vasylyev} {et~al.}(2022){Vasylyev}, {Filippenko}, {Vogl}, {Brink},
  {Brown}, {de Jaeger}, {Matheson}, {Gal-Yam}, {Mazzali}, {Modjaz}, {Patra},
  {Rowe}, {Smith}, {Van Dyk}, {Williamson}, {Yang}, {Zheng}, {deGraw}, {Fox},
  {Gates}, {Jennings}, \& {Rich}}]{2022ApJ...934..134V}
{Vasylyev}, S.~S., {Filippenko}, A.~V., {Vogl}, C., {et~al.} 2022, \apj, 934,
  134, \dodoi{10.3847/1538-4357/ac7220}

\bibitem[{{Vestrand} {et~al.}(2014){Vestrand}, {Wren}, {Panaitescu}, {Wozniak},
  {Davis}, {Palmer}, {Vianello}, {Omodei}, {Xiong}, {Briggs}, {Elphick},
  {Paciesas}, \& {Rosing}}]{2014Sci...343...38V}
{Vestrand}, W.~T., {Wren}, J.~A., {Panaitescu}, A., {et~al.} 2014, Science,
  343, 38, \dodoi{10.1126/science.1242316}

\bibitem[{{Wang} {et~al.}(2024){Wang}, {Wang}, {Wang}, {Zhou}, {Jin}, \&
  {Fan}}]{2024GCN.38055....1W}
{Wang}, Q.~L., {Wang}, Y., {Wang}, H., {et~al.} 2024, GRB Coordinates Network,
  38055, 1

\bibitem[{{Wren} {et~al.}(2010){Wren}, {Vestrand}, {Wozniak}, \&
  {Davis}}]{2010SPIE.7737E..23W}
{Wren}, J., {Vestrand}, W.~T., {Wozniak}, P., \& {Davis}, H. 2010, in Society
  of Photo-Optical Instrumentation Engineers (SPIE) Conference Series, Vol.
  7737, Observatory Operations: Strategies, Processes, and Systems III, ed.
  D.~R. {Silva}, A.~B. {Peck}, \& B.~T. {Soifer}, 773723,
  \dodoi{10.1117/12.859039}

\bibitem[{Zhou(2025)}]{hao_zhou_2025_14603585}
Zhou, H. 2025, {Clean Extraction for Swift/UVOT UV grism spectra}, 1.0,
  Zenodo, \dodoi{10.5281/zenodo.14603585}

\bibitem[{{Zhou} {et~al.}(2023){Zhou}, {Jin}, {Covino}, {Fan}, \&
  {Wei}}]{2023ApJS..268...65Z}
{Zhou}, H., {Jin}, Z.-P., {Covino}, S., {Fan}, Y.-Z., \& {Wei}, D.-M. 2023,
  \apjs, 268, 65, \dodoi{10.3847/1538-4365/acf20a}

\end{thebibliography}
\bibliographystyle{aasjournal}



\end{document}